\documentclass[preprint2]{aastex}
\begin{document}

\title{The Galactic Centre in the Far Infrared}

\author{M. Etxaluze}\affil{Harvard-Smithsonian Center for Astrophysics, 60 Garden Street, Cambridge, MA 02138, USA}

\author{Howard A. Smith}\affil{Harvard-Smithsonian Center for Astrophysics, 60 Garden Street, Cambridge, MA 02138, USA}

\author{V. Tolls }\affil{Harvard-Smithsonian Center for Astrophysics, 60 Garden Street, Cambridge, MA 02138, USA}

\author{A. A. Stark }\affil{Harvard-Smithsonian Center for Astrophysics, 60 Garden Street, Cambridge, MA 02138, USA}

\author{ E. Gonz\'alez-Alfonso}\affil{CfA and Universidad de Alcal\'a, Alcal\'a de Henares 28801, Spain}
\authoremail{metxaluz@cfa.harvard.edu}

\begin{abstract}
We analyse the far infrared dust emission from the Galactic Centre region, including the Circumnuclear Disk (CND) and other structures, using Herschel PACS and SPIRE photometric observations. These Herschel data are complemented by unpublished observations by the Infrared Space Observatory Long Wavelength Spectrometer (ISO LWS), which used parallel mode scans to obtain photometric images of the region with a larger beam than Herschel but with a complementary wavelength coverage and more frequent sampling with ten detectors observing at ten different wavelengths in the range from 46 to 180 $\mu$m, where the emission peaks. We also include data from the Midcourse Space Experiment (MSX) at 21.3 $\mu$m for completeness. We model the combined ISO LWS continuum plus Herschel PACS and SPIRE photometric data toward the central 2 pc in Sagittarius A$^*$ (SgrA$^*$), a region that includes the CND. We find that the FIR spectral energy distribution is best represented by a continuum that is the sum of three grey-body curves from dust at temperatures of 90, 44.5, and 23 K. We obtain temperature and molecular hydrogen column density maps of the region. We estimate the mass of the inner part of the CND to be $\sim5.0\times 10^4$ $\mbox{M}_{\odot}$, with luminosities: L$_{cavity}\sim2.2\times10^6$ $L_{\odot}$ and L$_{CND}\sim 1.5\times10^6$ $L_{\odot}$ in the central 2 pc radius around SgrA$^*$. We find from the Herschel and ISO data that the cold component of the dust dominates the total dust mass, with a contribution of $\sim3.2\times 10^4$ $\mbox{M}_{\odot}$; this important cold material had escaped the notice of earlier studies that relied on shorter wavelength observations. The hotter component disagrees with some earlier estimates, but is consistent with measured gas temperatures and with models that imply shock heating or turbulent effects are at work. We find that the dust grain sizes apparently change widely across the region, perhaps in response to the temperature variations, and we map that distribution. 
\end{abstract}


\section{Introduction}

The central $\sim$ 1 parsec of the galaxy is a low density central cavity that surrounds SgrA$^*$, the compact variable radio source located close to the dynamical centre of the Galaxy - a black hole of $\sim$4.5$\times$10$^6$M$_{\odot}$ at a distance of 8.3$\pm$0.4 Kpc \citep{Kerr86,Ghez08,Gillessen09}. The central cavity is characterized by emission of low ionisation atomic species. The cavity contains a cluster of Wolf-Rayet (WR) stars and O and B stars. Using far infrared (FIR) and submillimeter continuum data, \citet{Zylka95} estimated that the central cavity (radius $\sim$ 1 pc ) contains $\sim$400 $\mbox{M}_{\odot}$ of dust at T$_{kin}\sim$40 K, 4 $\mbox{M}_{\odot}$ at $\sim$170 K and $\sim$0.01 $\mbox{M}_{\odot}$ at $\sim$400 K.  The luminosities emitted by these three phases are $\sim 2\times 10^6$ $\mbox{L}_{\odot}$, $4\times10^6$ $\mbox{L}_{\odot}$, and $5\times10^5$ $\mbox{L}_{\odot}$ respectively. \citet{Latvakoski99} studied the region from the Kuiper Airborne Observatory (KAO) in the 30 $\mu$m band with a resolution of 8.5", and reported that  structures in the region have temperatures ranging from about 70 K to 100 K. They found the total luminosity of the components in the inner region (from their Table 1) to be $\sim$6$\times10^6$ $\mbox{L}_{\odot}$ and its mass $\sim$2000 $\mbox{M}_{\odot}$.

Surrounding the central cavity, extending from 1 pc to $\sim$ 5 pc, is the circumnuclear disk (CND). The CND has a mass of several thousand solar masses, and rotates with a velocity $\sim$ 100 Km$\cdot$s$^{-1}$. The neutral material in this ring is clumpy, dense (n(H$_2$) $\sim10^{3-7}$ cm$^{-3}$), and fragmented \citep{Genzel82,Genzel87} having a filling factor $\sim$10 $\%$ \citet{Zylka95,McGary01,Wright01}. There is a clumpy and filamentary structure known as the mini-spiral that connects the CND with the central ionised cavity, allowing ultraviolet radiation from the central ionised cavity to penetrate into the CND, heating and photoionising the gas \citep{Yusef01}.

The kinetic temperatures of the gas in the CND and central cavity ranges as high as 200 K - 300 K in CO(J = 7 - 6) \citep{Bradford05,Martin04} and $\sim$400 K in NH3 (6,6) \citep{Montero09,Herrnstein05}. The source of heating remains unclear. One explanation suggested is that the ionised central cavity is excited by the radiation from a central cluster of O and B stars, and that at the cavity edge the surrounding material is likely to be shock compressed by the expanding gas around the PDRs. Other possibilities include C-shock heating, perhaps by magnetic viscous heating, or possibly turbulent dissipation. \citet{Bradford05} suggest MHD (magneto hydrodynamics) shock heating with a shock velocity of order 20 - 30 Km$\cdot$s$^{-1}$, and a magnetic field 0.3 - 0.5 mG. If so, the material in the CND will dissipate its orbital energy within a few revolutions, and they suggest that its apparent longevity may result from refueling by infalling material from the surrounding clouds. Several authors suggest that the ionisation of the molecular clouds in the Galactic Centre is due to a large flux of cosmic rays heating the region \citep{Gusten81,Huttemeister93,Yusef07}.

\begin{figure*}[!ht]
\epsscale{2.05}
\plotone{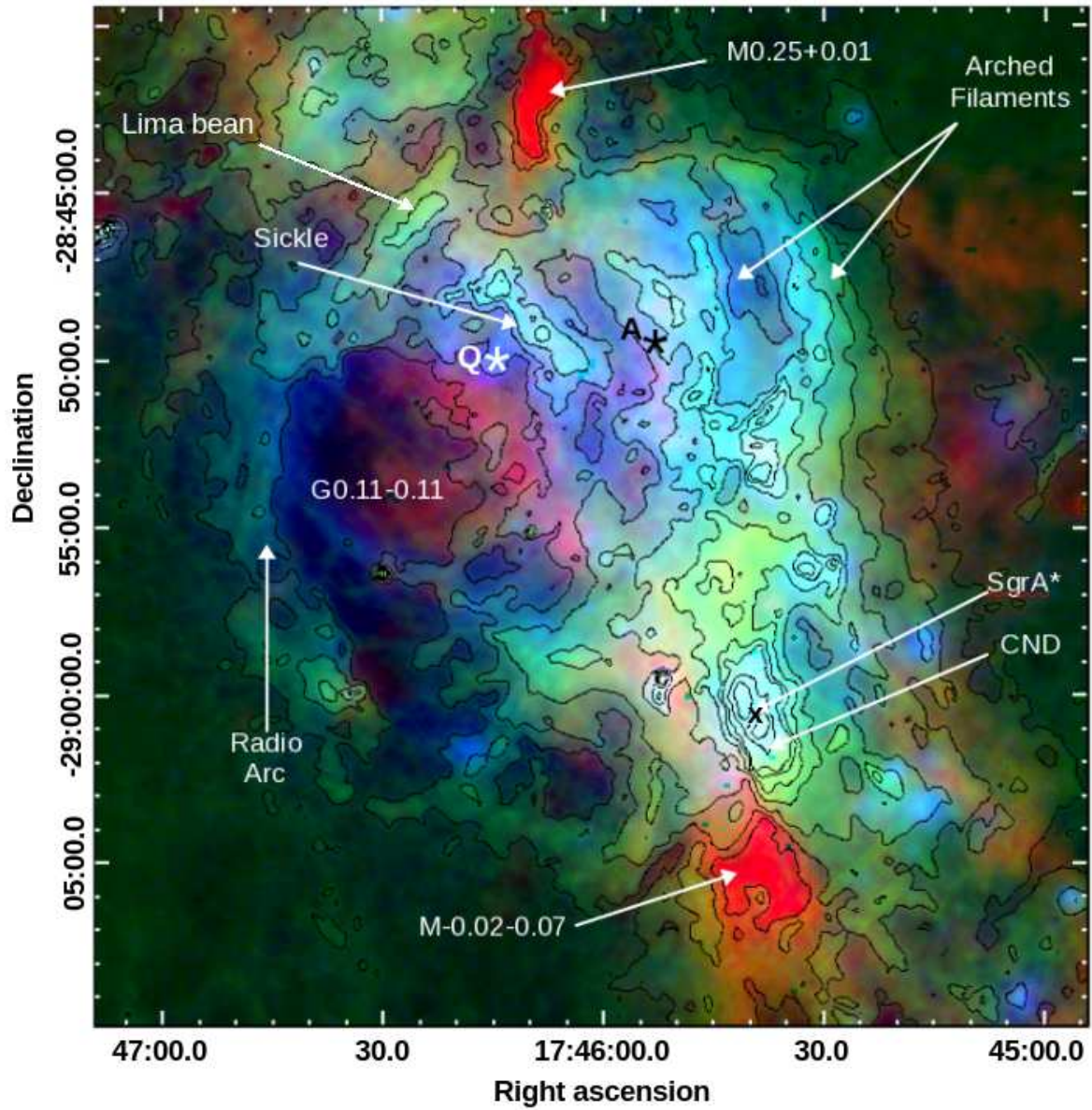}
\caption{Composite image of the SgrA$^*$ region; MSX-E at 21.3 $\mu$m (blue), 70 $\mu$m (green) and SPIRE 350 $\mu$m (red). The CND appears more prominent at 21.3 $\mu$m and 70 $\mu$m due to the lower angular resolution of the 350 $\mu$m observations. The black cross shows the position of Sgr A$^*$ at RA (2000)= 17h 45m 39s.77 Dec (2000)= -29$^o$ 00' 43.0". A and Q mark the position of the Arches and the Quintuplet clusters of O and WR stars, which provide enough photons to ionise and heat the Sickle and the Arched filaments. The position of the main molecular clouds (G0.11-0.11, M-0.02-0.07 and M0.25+0.01) across the Galactic Centre are also indicated. Each tic on the Declination axis is an arcminute, corresponding to 2.4 pc. Black contours represent fluxes 0.1, 0.45 1.0, 1.97, 2.89, 4.0, 5.0 and 14.0 Jy on the PACS 70 $\mu$m image.}
\label{fig:Composite}
\end{figure*}

The CND is surrounded by a number of dense molecular clouds and filamentary structures (see Figure~\ref{fig:Composite}). At a distance of $\sim$39 pc to the North East of SgrA$^*$ is the Radio Arc consisting of thermal and nonthermal structures aligned almost perpendicularly to the Galactic plane \citep{Yusef84}. The Radio Arc seems to connect to SgrA$^*$ via two concentrations of ionised and molecular gas: the Sickle and the Arched Filaments. Two clusters of WR and O-type stars, the Quintuplet and the Arches Cluster, are responsible for the ionisation of the Sickle and the Arched filaments, respectively. The Arches cluster is slightly in the foreground, about 20 pc from the Arched filaments, and irradiates the several filaments approximately uniformly \citep{Lang01}.

The MSX image at 21.3 $\mu$m and the PACS image at 70 $\mu$m trace a shell surrounding the Arches and Quintuplet clusters. The shell goes through the Arched Filaments, the Lima bean and passes through the south of the G0.11-0.11 molecular cloud forming a ring of $\sim$44 pc in diametre (Figure~\ref{fig:Composite}). \citet{Bally10} suggest that the radiation and the stellar winds from the Arches and the Quintuplet clusters and the black hole in SgrA$^*$ may be responsible for the formation of this shell, which can clearly be seen at the Spitzer 24 $\mu$m image \citep{Yusef09}, and which has also been reported by \citet{Moneti01}.

The objective of this paper is to further constrain the physical conditions in the central 2 parsecs of the Galactic Centre, and to help clarify the excitation conditions. We use archival Herschel PACS and SPIRE photometric observations of the region to determine the dust properties more accurately; many earlier data sets at shorter wavelengths \citep{Latvakoski99} would have been insensitive to a cold dust component. We also incorporate previously unpublished infrared images from the Infrared Space Observatory Long Wavelength Spectrometer (ISO-LWS), operating in parallel mode; its grating sampling allows us to obtain a more precise analysis of the character and structure of dust in the central region and the CND.

\section{Observations}

Herschel PACS \citep{Poglitsch01} data at 70 $\mu$m and 160 $\mu$m, and SPIRE \citep{Griffin06} data at 250 $\mu$m, 350 $\mu$m and 500 $\mu$m, were used to measure the dust temperature and the column density distribution across the Galactic Centre region. The basic data were obtained from the public Herschel archive, and were reprocessed using HIPE, the Herschel Interactive Processing Environment. The data were collected in parallel mode and are still rather preliminary in absolute accuracy, suffering from calibration errors.  PACS fluxes off the bright galactic plane are negative, indicating that the pipeline corrections are inaccurate; moreover there are indications that in this extremely bright region some non-linear detector responses are uncorrected. We do not apply any across-the-board offsets to get positive flux values. As a result, the derivative images we discuss below are reliable over most but not all of the area, but the SEDs we fit to the warm central region (see Section 4) are accurate. Images are made in Jy/pix at 2.0$''$/pixel and 3.0$''$/pixel for PACS 70 $\mu$m and 160 $\mu$m, respectively, and in Jy/beam for SPIRE data at 250, 350 and 500 $\mu$m, with beam areas respectively of 395.0$''^2$ , 740.0$''^2$ and 1571.0$''^2$. Figure~\ref{fig:Composite} covers an area of  $28.46\times29.4$ arcmin$^2$ centred at RA(2000)= 17h 45m 39.77, Dec(2000)= -29$^0$ 00' 43.0". For a distance of 8.5 Kpc, the covered area is $\sim$$70.4\times72.7$ pc$^2$. 

\begin{table}[!ht]
\begin{center}
\footnotesize{\begin{tabular}{cccccc}
\tableline
\tableline
\multicolumn{6}{c}{ }\\
&\multicolumn{2}{c}{PACS}&\multicolumn{3}{c}{SPIRE}\\
$\lambda(\mu\mbox{m})$&70.0&160.0&250.0&350.0&500.0\\
\tableline
\multicolumn{6}{c}{ }\\
pixel ($''$)&2.0&3.0&6.0&10.0&14.0\\
FWHM ($''$)&5.9&11.6&18.5&25.3&36.9\\
FWHM (pc) &0.24&0.48&0.76&1.04&1.52\\
\tableline
\multicolumn{6}{c}{ }\\
\end{tabular}}
\caption{PACS and SPIRE pixel and beam sizes}
\end{center}
\label{tab:Herschel_parameters}
\end{table}

The image is color-stretched to show more of the faint or cool structures, which are labeled. The most obvious places where the photometry issues make analyses difficult are in the coldest, dark regions, for example in G0.11-0.11.  We white-out the corresponding area in the derivative column density and spectral index images because of the uncertain correction to the negative fluxes reported by the current Herschel pipeline. 

The MSX (Midcourse Space Experiment) surveyed the Galactic Plane at 6 bands from 4.2 to 26 $\mu$m at a spatial resolution of 18" \citep{Egan99}. The MSX data can also be used to study the morphology, dynamics and physics of the ISM providing the thermal emission from dust grains at mid infrared wavelengths \citep{Cohen99}. We used the MSX Band E data at 21.3 $\mu$m to trace the warm dust across the Galactic Centre. 

We also report in this paper previously unpublished observations of Sgr A$^*$, taken by the Long Wavelength Spectrometer (spectral range 46-180 $\mu$m) on the Infrared Space Observatory satellite in parallel mode. ISO was equipped with two spectrometers: SWS and LWS, a camera: ISOCAM, and an imaging photopolarimeter: ISOPHOT to cover wavelengths from 2.5 to 240 $\mu$m with spatial resolutions of 1.5", at the shortest wavelengths, and 85"-90" at the longer wavelengths. Two instruments, LWS and ISOCAM, were able to operate simultaneously in parallel mode. More than 100,000 individual pointings were made in parallel mode in over 17,000 individual observations with a total sky coverage of about 1$\%$. Key to our study, the LWS operated with ten wavelengths in the range 46-180 $\mu$m \citep{Lim99}, each with bandwidths of about 0.3 $\mu$m for the five short wavelength detectors (SW) and 0.6 $\mu$m for the five long wavelength detectors (LW). Parallel mode maps are generated by combining different raster scans. Several data reduction tools were developed in the IDL programming language and these form the LWS parallel interactive analysis package that can be found at: 
{\footnotesize http://jackal.bnsc.rl.ac.uk/iso/lws/software/software.html}

Now that the Herschel observations have obtained longer wavelength coverage, out to 500 $\mu$m, the earlier ISO data can reliably be combined with them to prepare a more comprehensive picture. In particular, the spectral information obtained with ISO-LWS allows us to determine the peak wavelength and flux of the dust emission with much greater precision than is possible with broad-band photometry; this in turn allows us to specify the conditions of excitation much more accurately.

\section{Dust grain properties; temperature and optical depth maps}
Herschel PACS and SPIRE photometry have a much higher spatial resolution in the FIR than any previous mission. When analysed, the maps provide a much improved spatial distribution of the fluxes, temperature, spectral index, and H$_2$ column density across the Galactic Centre region. 
The MSX 21.3 $\mu$m images provide an important measure of the warmer material in the region. Figure~\ref{fig:Composite} shows the Galactic Centre as seen by combined images of MSX-E 21.3 $\mu$m (blue), PACS 70 $\mu$m (green) and SPIRE 350 $\mu$m (red). The arched filaments, the Radio Arc and Sgr A$^*$ are clearly visible at 21.3 $\mu$m and 70 $\mu$m, while SPIRE 350 $\mu$m traces the very cold and dense molecular clouds. Herschel resolves the central cavity and the surrounding CND at 70, 160 and 250 $\mu$m; the lower SPIRE 350 and 500 $\mu$m resolutions (FWHM= 25.3"and 36.9" respectively) are not quite able to resolve the cavity. 

\begin{figure}[!hb]
\epsscale{1.00}
\plotone{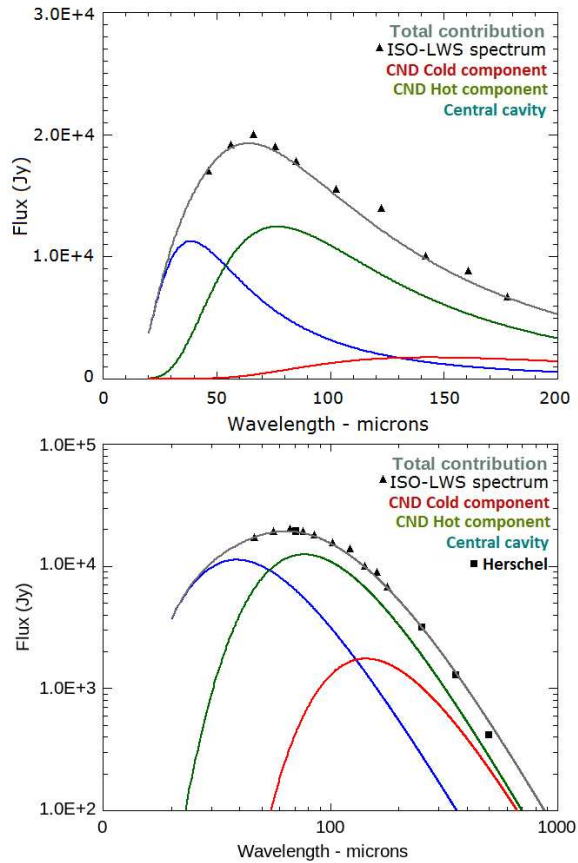}
\caption{(Top) Integrated fluxes in the ISO-LWS beam for each detector (black triangles) centred at the position of SgrA$^*$. The continuum is fitted by the sum of three different dust components: (blue) a very hot (T$_{dust}= 90$ K) with a very low density component (n$_{H_2} \sim 2.8\times10^3$ cm$^{-3}$) arising from the central cavity, and two different dust component coming out from the inner part of the CND: (green) a hot component (T$_{dust}= 45$ K) with a density (n$_{H_2} \sim 1.19\times10^4$ cm$^{-3}$) and (red) a cold component (T$_{dust}= 23.5$ K) with very high density (n$_{H_2} \sim 2.41\times10^4$ cm$^{-3}$). The grey line represents the total contribution. (Bottom) The continuum extended to longer wavelengths. The black squares represent the Herschel fluxes integrated on the ISO-LWS beam at each detector: PACS at 70 $\mu$m, and SPIRE at 250, 350 and 500 $\mu$m }
\label{fig:Total_continuum}
\end{figure}

The observed continuum emission from the Galactic Centre region peaks at about 66 $\mu$m, in the range of ISO-LWS detector SW3 (Figure~\ref{fig:Total_continuum}). As shown in Figure~\ref{fig:Total_continuum}, the shape of the spectral energy distribution (SED) cannot be fit by a single grey body curve. In the ensuing discussion, we present our procedure for decomposing the shape, which depends on the temperature, the number of grains along the line of sight, and their emissivity and spectral index, $\beta$. At the longest wavelengths, the grain properties determine the shape of the SED. We assume a grain size of $a\sim0.1$ $\mu$m. The temperature and the spectral index, $\beta$, distribution across the Galactic Centre was then obtained by fitting the PACS and SPIRE wavelengths with a grey body curve: 
   
\begin{equation}
F_{\nu}=\Omega\times B_{\nu}\left( T \right)\times\left( 1-e^{-\tau_{\nu}} \right) 
\label{eq:greybody}
\end{equation}
where $\Omega$ is the solid angle, T is the temperature and $\tau_{\nu}$ is the optical depth for a frequency $\nu$, given as:
\begin{equation}
\tau_{\nu}=\tau_0 \left(\frac{\nu}{\nu_0}\right)^{\beta} 
\label{eq:tau}
\end{equation}  
$\tau_0$ is the bibitem optical depth at $\nu_0$, the bibitem frequency, which was chosen to be c/100 $\mu$m. The shape of the curve is only mildly sensitive to the choice of $\nu_0$ \citep{Lewis00}. 

The SED for each pixel in the Herschel image was fitted after smoothing the images and adjusting the pixel sizes uniformly to the resolution of the SPIRE 500 $\mu$m channel band. The H$_2$ column density was then derived assuming a gas-to-dust-ratio of 100. We did not include the PACS 160 $\mu$m image data with the other four Herschel bands in the calculation of the temperature and the column density maps. 

At the brightest regions, the flux values from this detector did not fit any physical SED, perhaps due to a poor correction of low level stripes in the data and/or errors on the background subtraction. Moreover, in the low intensity regions the detector map has negative values possibly due to uncorrected gain adjustments \citep{Bernard10}. Its omission will not affect the conclusions significantly; subsequent pipeline reductions will hopefully address these issue.
 
\begin{figure}[!ht]
\epsscale{1.00}
\plotone{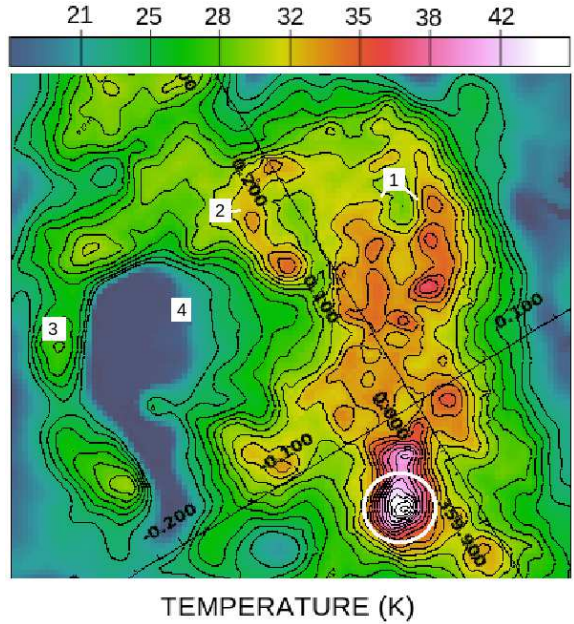}
\caption{Temperature map calculated with the PACS 70 $\mu$m and SPIRE 250, 350 and 500 $\mu$m. The ISO-LWS beam size is shown with a white circle centred at the position of SgrA$^*$. The black axis represents longitude (from top to bottom) and latitude (from left to right) in Galactic coordinates. (1) shows the position of the Arched filaments, (2) the Sickle, (3) the Radio Arc and (4) points at the position of the molecular cloud G0.11-0.11. The size of the image is $\sim24\times 24$ arcmin$^2$ ($\sim59.4\times 59.4$ pc$^2$, for a distance of 8.5 Kpc).}
\label{fig:Temperature}
\end{figure}

\begin{figure}[!ht]
\epsscale{1.00}
\plotone{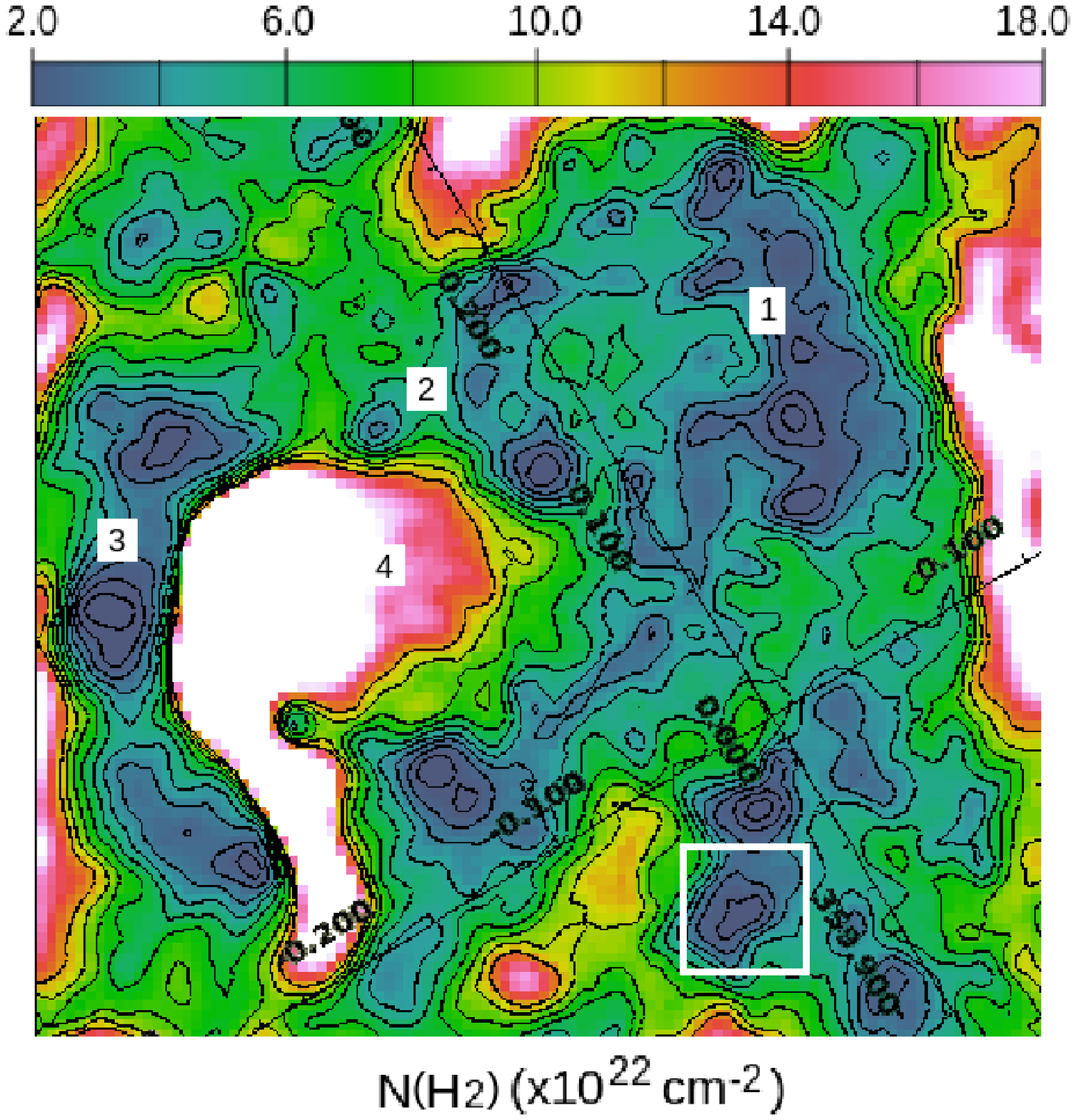}
\caption{H$_2$ column density map calculated with the PACS 70 $\mu$m and SPIRE 250, 350 and 500 $\mu$m. The white square represents the area covered by Figure~\ref{fig:MSXE_CND}. The scale was set to a range of (2 - 18)$\times$10$^{22}$ cm$^{-2}$. G0.11-0.11 molecular cloud (4) shows very high H$_2$ column density values N(H$_2$)$\sim$(30 - 80)$\times$10$^{22}$ cm$^{-2}$. The whole molecular cloud and the surrounding areas appear like a white bulk because the scale was limited to values  N(H$_2$)$<$18$\times$10$^{22}$ cm$^{-2}$ for a better definition of the lowest column density areas on the map. The size of the image is $\sim24\times 24$ arcmin$^2$ ($\sim59.4\times 59.4$ pc$^2$, for a distance of 8.5 Kpc).} 
\label{fig:column_density}
\end{figure}

Figures~\ref{fig:Temperature}, ~\ref{fig:column_density} and ~\ref{fig:Emissivity} show the resultant temperature, molecular hydrogen column density and $\beta$ maps, respectively; north is up in all cases. The white circle on the temperature map, centred at SgrA$^*$, shows the average size of the ISO-LWS beam which covers, approximately, the central cavity and the inner 1 pc of the CND, for a distance to the region of 8.5 Kpc. Figure~\ref{fig:MSXE_CND} shows the PACS 160 $\mu$m map; although the absolute calibration is defective, as noted, its higher resolution provides an improved picture of the CND than does either the temperature or column density maps, which have been reduced to the SPIRE 500 $\mu$m beam size and are therefore unable to resolve the CND and cavity structure. 

\subsection{The temperature map}
The temperature map (Figure~\ref{fig:Temperature}) shows the most distinguishing details of the various structures in this complex region. As expected, due to the strong UV radiation and the stellar winds from the WR and the OB stellar clusters placed at the central cavity, as well as the electron heating due to magneto-hydrodynamic turbulences that take place in accretion states around the central black hole \citep{Liu11}, SgrA$^*$ shows the highest temperature values , where T$_{dust}\gtrsim$ 60 K.

 This map beautifully resolves the Arched Filaments and Sickle with dust temperatures T$_{dust}\sim35$ K, and the Radio Arc, with significantly lower temperatures than the Arched Filaments: T$_{dust}\sim$28 K. As expected, the coldest region in the temperature map occurs in the giant molecular cloud G0.11-0.11, whose temperatures are T$_{dust}\lesssim 18$ K. As we discuss in more detail below and in Figure~\ref{fig:GC_ISO_LWS}, these results are in agreement with the ISO-LWS images, in which the Arches (for example) fade away as the wavelength changes from 46 $\mu$m to 178 $\mu$m. \citet{Morris95} used the KAO at 50 and 90 $\mu$m to map this region.  They infer a dust temperature for the Arches of 45-55 K using a $\beta$ = 1, noticeably warmer and inconsistent with our result, in some part because $\beta$ is actually closer to 2.

\citet{Colgan96} reported that the more eastern filament (called E2) might even be higher than 70 K.  The new Herschel - ISO results show that there is considerable knotty structure along each of the filaments and elsewhere across the region, but that average dust temperatures are more accurately closer to 35 K.
\subsection{The column density map}
The column density map (Figure~\ref{fig:column_density}) reveals the presence of numerous dense features most of which have been previously reported. 
\citet{Bally10} discuss many of them based on their SHARCII 0.35mm and BOLOBAM 1.1 mm continuum images with a
resolution of 9" and 33' respectively.

 Figure~\ref{fig:column_density} shows only about a factor of ten variations across most of the region, with few distinct structures being defined by their high column densities. The exception is the well defined, very dark cloud G0.11-0.11 with very low dust temperature, between 12 K and 18 K, and whose peak column densities exceed N(H$_2$)$\gtrsim3.0\times10^{23}$cm$^{-3}$. By contrast, the total luminosity varies like T$^{4+\beta}$ across the region, in which $\beta$ changes from about 1 to 3, and the dust surface brightness varies by a factor of about 300 even in the less dense structures. The column density map closely anti-correlates with the temperature map, with denser regions being colder.  

\begin{figure}[!ht]
\epsscale{1.00}
\plotone{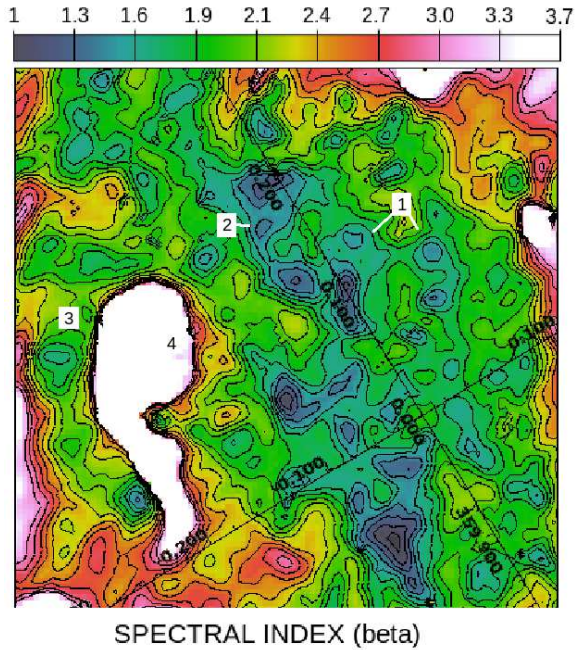}
\caption{Spectral index map calculated with the PACS 70 $\mu$m and SPIRE 250, 350 and 500 $\mu$m. (1) shows the position of the Arched filaments, (2) the Sickle, (3) the Radio Arc and (4) points at the position of the molecular cloud G0.11-0.11. The size of the image is $\sim24\times 24$ arcmin$^2$ ($\sim59.4\times 59.4$ pc$^2$, for a distance of 8.5 Kpc).} 
\label{fig:Emissivity}
\end{figure}

As a result of the low fluxes, as noted earlier, the PACS photometry in the area of G0.11-0.11 has uncorrected negative values, and the density and $\beta$ values are approximately only. In Figures~\ref{fig:column_density} and~\ref{fig:Emissivity}, the scale has been intentionally stretched to highlight the less prominent structures at the expense of saturating the G0.11-0.11 area).
The cloud itself is more accurately defined by the SPIRE emission (see Figure~\ref{fig:Composite}, in red), and indeed the images reveal considerable knots and substructure across the source. The southerly extension to the source, visible in the three derived maps, is an artifact of the negative photometry in the low flux region; SPIRE emission clearly shows that the low flux is due to an absence of material rather than cold high density dust. We can nevertheless estimate a lower limit to the total mass of G0.11-0.11 as 1$\times10^{6}$ M$_{\odot}$.

\begin{figure}[!ht]
\epsscale{1.00}
\plotone{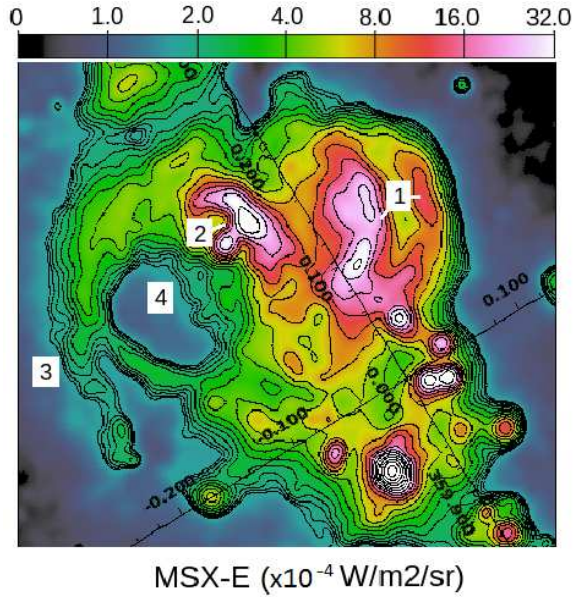}
\caption{MSX-E image at 21.3 $\mu$m. (1) shows the position of the Arched filaments, (2) the Sickle, (3) the Radio Arc and (4) points out the position of the molecular cloud G0.11-0.11. The size of the image is $\sim24\times 24$ arcmin$^2$ ($\sim59.4\times 59.4$ pc$^2$, for a distance of 8.5 Kpc).}
\label{fig:MSXE}
\end{figure}

\subsection{The spectral index map}
The map for $\beta$ is shown in Figure~\ref{fig:Emissivity}; it is also anti-correlated with the temperature map (Figure~\ref{fig:Temperature}). SgrA$^*$, for example, the hottest region, shows the lowest values of $\beta$$\sim$ 1.0-1.3. The Arched Filaments and Sickle show values in the range $\beta$$\sim$1.3-1.8 and the range of $\beta$ values in the Radio Arc is $\beta$$\sim$1.7-2.2. The G0.11-0.11 molecular cloud shows extremely large values of $\beta\sim$3.5. This general trend is expected; dust grains in cold clouds tend to be larger in size and therefore have larger $\beta$ than grains in warm clouds \citep{Butler09,Hirashita07,Dupac03}. The strong and rapid spatial variations in the Galactic Centre, coupled with the Herschel maps, provide one of the first clear looks at how this feature of grains manifests itself in a complex region. 

Since the total luminosity varies roughly like T$^{4+\beta}$, this parameter also helps sort out the excitation and energetics of the various structures. The anticorrelation between temperature and density applies to most of the other dense regions of the maps as well, which, being dominated by dust properties, are coldest where they are most opaque (Figure~\ref{fig:column_density}); for example through the Arched Filaments, Sickle, the Radio Arc and SgrA$^*$. 

\subsection{The distribution of the very small grains (VSG)}
The O and WR stars from the Arches and the Quintuplet clusters (Figure~\ref{fig:Composite}) provide enough photons to ionise and heat Sickle and the Arched filaments \citep{Cotera96} and also generate strong winds that could contribute to the depletion of the largest dust grains whose population is traced by the N(H$_2)$ map: The largest dust grains can be destroyed due to the strong winds and the UV radiation from the massive O and WR stars from the Arches and Quintuplet clusters. 

\begin{figure}[!ht]
\epsscale{1.00}
\plotone{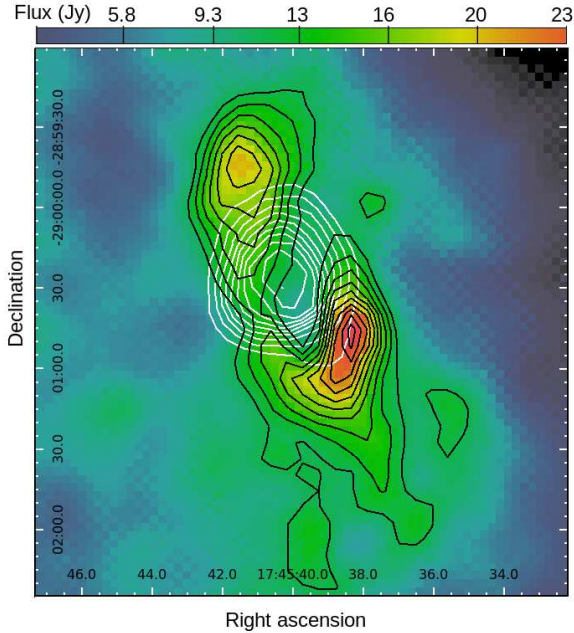}
\caption{SgrA$^*$ image from PACS 160 $\mu$m. The map resolves the CND at 160 $\mu$m (black contours) surrounding the central cavity, which is well-traced by the MSX-E contours (white) at 21.3 $\mu$m. The MSX-E intensity peaks in the centre of the cavity at 3.89$\times10^{-3}$ W/cm$^2$/sr decreasing outwards to 1.0$\times10^{-3}$ W/cm$^2$/sr. The size of the map is $\sim3.3\times 3.4$ arcmin$^2 (\sim8.2\times 8.4$ pc$^2$, for a distance of 8.5 Kpc).}
\label{fig:MSXE_CND}
\end{figure}

The MSX-E map at 21.3 $\mu$m (Figure~\ref{fig:MSXE}) can be used to trace the population of very small grains (VSG) throughout the Galactic Centre region (this band is more reliable than the features observed at shorter infrared wavelengths which can be affected by extinction). VSGs are small (with sizes of $\sim0.01$ $\mu$m), and absorption of one UV photon \citep{Pomares09} can increase their temperature up to 80 K, prompting them to re-emit at these shorter wavelengths ($\lambda\sim24$ $\mu$m) \citep{Desert90,Boulanger85}. The MSX-E map correlates well with the temperature map, showing that the population of the small grains is distributed through the regions where the population of the largest grains has been depleted, i.e. the Sickle and the Arched filaments. This effect can also be observed in SgrA$^*$: PACS 160 $\mu$m, Figure~\ref{fig:MSXE_CND} , shows very low emission at this far-infrared wavelength in the central cavity of SgrA$^*$, but which is surrounded by bright emission from the CND. The emission in the inner parsec, as traced by MSX-E at 21.3 $\mu$m, appears confined to the central cavity where the O and B stars from the central cluster produce a strong radiation field, and where stellar winds heat up the small grains and deplete large grains. Within the central 2 parsecs, the largest grains are found remaining in the CND itself, where they are shielded from the central strong UV radiation; their emission is traced by Herschel PACS+SPIRE at far infrared wavelengths $\lambda > 60$ $\mu$m.

\section{The Temperature Components of the CND: Fitting the ISO LWS continuum}

ISO-LWS obtained images of the central region of the galaxy operating in its parallel mode, one image simultaneously from each of the ten ISO-LWS detectors, with the spectrometer's grating set at a fixed position near the centre of its scan range. The ISO LWS parallel mode maps provide us with complementary wavelength coverage just where it is most needed, at the peak of the dust emission, in the range of 46 to 180 $\mu$m that can be used to defined the SED of SgrA$^*$ at far infrared wavelengths. Unfortunately, with a beam of $\sim 85"$ and a pixel size of 42", these ISO maps are unable to resolve as many structures as do higher resolution Herschel PACS and SPIRE data.

\begin{figure*}[!ht]
\epsscale{2.00}
\plotone{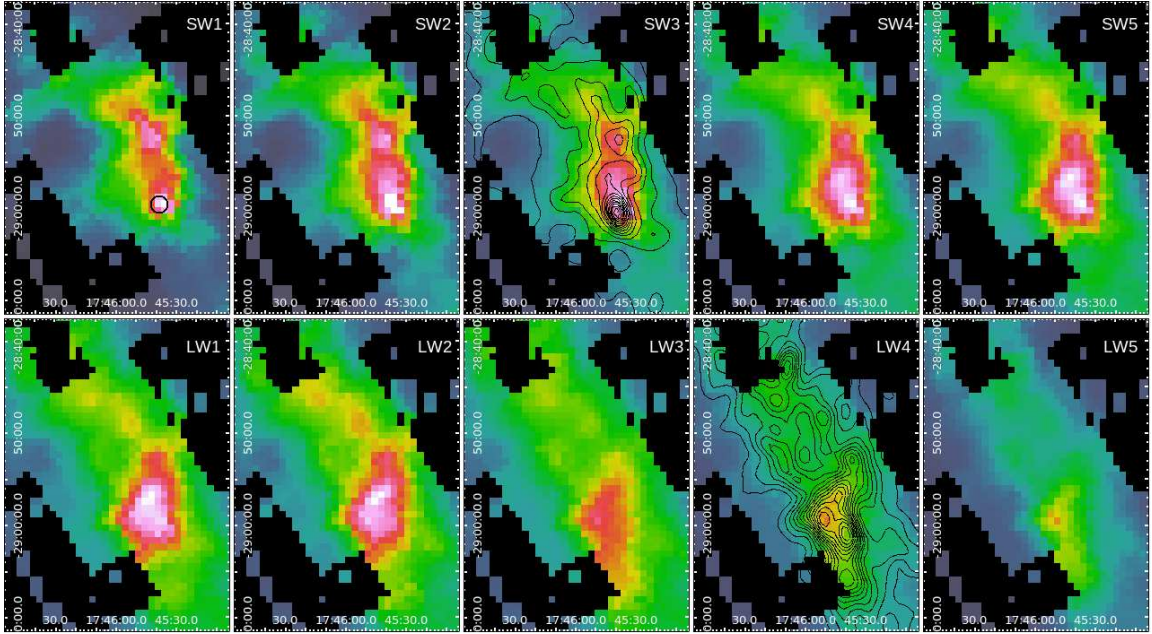}
\caption{Galactic Centre observed at each ISO-LWS parallel mode wavelength: 46.222 $\mu$m (SW1), 56.2033 $\mu$m (SW2), 66.1173 $\mu$m (SW3), 75.6989 $\mu$m (SW4), 84.7977 $\mu$m (SW5), 102.425 $\mu$m (LW1), 122.218 $\mu$m (LW2), 141.809 $\mu$m (LW3), 160.554 $\mu$m (LW4) and 177.971 $\mu$m (LW5). Black contours on SW3 are PACS 70 $\mu$m convolved with the SPIRE 500 $\mu$m beam size (36.9") and the black contours on LW4 correspond to PACS 160 $\mu$m emission with a beam size of 36.9". The black circle on the SW1 map represents the ISO-LWS beam ($\sim 3.5$ pc) centred at the position of SgrA$^*$. }
\label{fig:GC_ISO_LWS}
\end{figure*}

Figure~\ref{fig:GC_ISO_LWS} shows the resultant Galactic Centre images in each of the ten bands. The images are centred at RA (2000)= 17h 45m 56s.38, Dec (2000)= -28$^0$ 55' 15.24" an cover and area of $\sim$ 22.8$\times$29.6 arcmin$^2$ ($\sim$ 59$\times$77 pc$^2$). In order to obtain accurate photometry from these images, an extended source flux correction factor is required, as described in the ISO Users Manual. Fortunately we also had pointed observations with which to confirm the correction procedure (see \citep{White11}, for a description of the ISO pointed photometric and spectroscopic observations). The observed flux in Jy is:
\begin{equation}
S_{\nu}[Jy]= F_{\nu}\times f / (\Omega \times 10^6)[MJy\cdot sr^{-1}] 
\label{eq:ISOcorrection}
\end{equation}

and the extended source correction factors ($f$) and the solid angle $\Omega$ are given in Table 2. Besides, \citet{Casassus08} reported that the ISO LW1 at 102.42 $\mu$m intensities were 27$\%$ higher than IRAS 100 $\mu$m for intensities bellow 1000 MJy$\cdot$sr$^{-1}$ and 40$\%$ higher for intensities above 1000 MJy$\cdot$sr$^{-1}$. This was in agreement with \citet{Chan01}. This offset on the ISO parallel mode data was calculated by comparing the fluxes at different position along the Galactic Centre observed in primary mode with the fluxes observed in parallel at the same position for the ten ISO LWS detectors. The correction factors, $\mathcal{Q}$($Flux_{parallel}/Flux_{primary}$), are also given on Table 2.

\begin{table}[!ht]
\begin{center}
\footnotesize{\begin{tabular}{ccccc}
\tableline
\tableline
\multicolumn{5}{c}{ }\\
{\bf Detector} & {\bf $\lambda$($\mu$m)}&{\bf  $f$}&{\bf $\Omega\times 10^6$ [sr]}&{\bf $\mathcal{Q}$}\\
\tableline
\multicolumn{5}{c}{ }\\
SW1 & 46.22&0.8704&0.1140&1.082\\
SW2 & 56.20&0.8677&0.1321&0.899\\
SW3 & 66.11&0.8421&0.1410&0.889\\
SW4 & 75.69&0.7334&0.1223&0.823\\
SW5 & 84.79&0.6878&0.1163&0.878\\
LW1 &102.42&0.6758&0.1111&0.930\\
LW2 & 122.21&0.6734&0.1128&0.823\\
LW3 & 141.80&0.6035&0.0935&0.948\\
LW4 & 160.55&0.5411&0.0904&0.992\\
LW5 & 177.97&0.4596&0.0833&1.004\\
\tableline
\multicolumn{5}{c}{ }\\
\end{tabular}}
\caption{Table of extended source correction ($f$) and effective solid angle of the beam ($\Omega$) for the
different LWS detectors.}
\end{center}
\label{tab:ISO_LWS_detectors}
\end{table}

Despite the correction factors were applied, the LW2 fluxes seem to be still offset in our data by 10$\%$ above the 
continuum level. Due to this problem, the LW2 flux was omitted when fitting the ISO-LWS SED in Figure~\ref{fig:Total_continuum} .

SgrA$^*$ is very bright at the shortest wavelengths, and clearly seen in Figure~\ref{fig:GC_ISO_LWS}. The SW detectors (46.2 - 84.79 $\mu$m) also show the northern part of the Radio Arc. Detectors SW4 75.69 $\mu$m, SW5 84.79 $\mu$m, LW1 102.42 $\mu$m and LW2 122.2 $\mu$m show SgrA$^*$ as a single object about $\sim$4 pc in radius that includes SgrA$^*$ and extends to M-0.02-0.07 molecular cloud (Figure~\ref{fig:Composite}) in the south. At 160.55 $\mu$m (LW4) and 177.97 $\mu$m (LW5), the most conspicuous feature is the cold and dense molecular cloud M-0.02-0.07 (see also Figure~\ref{fig:Composite}).  

The spectral energy distribution of Sgr $A^*$ within the $\sim$85$''$ ISO-LWS beam is shown in Figure~\ref{fig:Total_continuum}. Each data point on Figure~\ref{fig:Total_continuum} represents the integrated flux on the ISO-LWS beam centred at the position of SgrA$^*$ at each wavelength. We note that at a distance to the region of about 8.5 Kpc, the ISO-LWS  beam size of 3.5 pc includes the central cavity and the inner $\sim$0.8 pc of the CND, not quite large enough to completely include the full CND observed in the Herschel maps. In order to model the ISO-LWS continuum spectrum across the wavelength range $\sim$40-200 $\mu$m, we assumed typical dust properties: dust grains of 0.1 $\mu$m in radius and a bulk density of 2.5 (g$\cdot$cm$^{-3}$). These values are characteristic of silicate grains that contain H$_2$O inside their structure (e.g. \citet{Pollack94}Pollack et al. 1994). The dust opacity is defined as:
\begin{equation}
\kappa_{\lambda}\cdot\rho=\frac{\kappa_{const}}{\lambda^{\beta}}
\label{eq:opacity}
\end{equation}
where $\kappa_{\lambda}$(cm$^2\cdot$g$^{-1}$) is the dust opacity, $\rho$(g$\cdot$cm$^{-3}$) is the bulk density of the grains and $\kappa_{const}$ is a proportionality constant. For $\lambda=$100 $\mu$m and $\kappa_{const}$= 0.8 the dust opacity in the central cavity is $\kappa_{100\mu m}\sim$ 92 cm$^2\cdot$g$^{-1}$,  characteristic of dust grains with thin ice mantles, while the opacities at the CND are $\sim$160 and 200 cm$^2$/g for the hot and the cold component, 
respectively. These values are typical of dust grains with thick ice layers \citep{Ossenkopf94}.

The best model that we found able to reproduce the LWS spectral continuum consists of three-temperature components: an SED arising from the central cavity of r=1 pc in radius, and a combination of a warm plus cold SED arising from the inner 0.86 pc of the CND. Figure~\ref{fig:column_density} shows the steep gradient in column density at the position of SgrA$^*$, from values of NH$_2\sim3\times10^{22}$ cm$^{-2}$ at the centre of the cavity to higher values $\sim1\times10^{23}$ cm$^{-2}$ at the CND. The temperature map (Figure~\ref{fig:Temperature}) likewise shows the largest values just on the centre , T$_{dust}\gtrsim$ 60 K, surrounded by colder dust. The SED from the cavity can be fitted by a grey-body curve defined by a dust temperature of T$_{dust}$= 90 K, $\beta$=1.23 and a low column density of N(H$_2$)= $8.73\times 10^{21}$ cm$^{-2}$.

As noted, the SED of the CND requires two components in order to fit the whole ISO-LWS continuum: a hot component of T$_{dust}$= 44.5 K, $\beta$=1.35 and a column density of N(H$_2$)= $3.18\times 10^{22}$ cm$^{-2}$.  Significantly, a cold component also has to be introduced in order to fit the longest wavelengths of the ISO-LWS data. The cold component can be fitted by a T$_{dust}$= 23.5 K modified blackbody with $\beta$=1.40 and a higher column density N(H$_2$)= $6.39\times 10^{22}$ cm$^{-2}$. This massive new cold component, previously unreported in observations done only at shorter wavelengths, is reconfirmed by the longer wavelength Herschel photometry. This simple, three component model reproduces very well the observed continuum emission at the ISO-LWS spectrum (Figure~\ref{fig:Total_continuum}) and the SPIRE and PACS 70 $\mu$m data. The luminosity of each component is: L$_{cavity}\sim 2.2\times10^6$ $L_{\odot}$, L$_{CND}^{hot}\sim 1.4\times10^6$ $L_{\odot}$ and L$_{CND}^{cold}\sim 1.0\times10^5$ $L_{\odot}$, and the total gas mass: M$_{cavity}\sim 1.7\times10^3$ $M_{\odot}$, M$_{CND}^{hot}\sim 1.56\times10^4$ $M_{\odot}$ and M$_{CND}^{cold}\sim 3.15\times10^4$ $M_{\odot}$. Table \ref{tab:Paramenters} lists the parameters from the fit to the ISO-LWS continuum.

\begin{table}[!h]
\begin{center}
\footnotesize{\begin{tabular}{cccc}
\tableline
\tableline
\multicolumn{4}{c}{ }\\
& {\bf Cavity} &{\bf CND$_{hot}$} & {\bf CND$_{cold}$} \\
\tableline
\multicolumn{4}{c}{ }\\
$radius$$\left(pc\right)$&1.0&0.86&0.86\\
$\beta$&1.23&1.35&1.40\\
$\kappa_{const.}$&0.8&0.8&0.8\\
$\kappa_{100\mu m}$&92.28 &160.38 &200.\\
$T_{dust}$$\left(K\right)$& 90.0& 44.5&23.5\\
$\rho_{dust}$$\left(cm^{-3}\right)$&9.45$\times$10$^{-23}$&4.0$\times$10$^{-22}$&8.05$\times$10$^{-22}$\\
$\tau_{100\mu m}$&5$\times$10$^{-3}$&3$\times$10$^{-2}$&8$\times$10$^{-2}$\\
$N_{H_2}$$\left(cm^{-2}\right)$&8.73$\times$10$^{21}$&3.18$\times$10$^{22}$&6.39$\times$10$^{22}$\\
$n_{H_2}$$\left(cm^{-3}\right)$&2.8$\times$10$^{3}$&1.19$\times$10$^4$&2.41$\times$10$^4$\\
$\chi_{dust}$&0.01&0.01&0.01\\
$L\left(L_{\odot}\right)$&$2.2\times10^6$&$1.4\times10^6$&$1.0\times10^5$\\
$M\left(M_{\odot}\right)$&$1.7\times10^3$&$1.56\times10^4$&$3.15\times10^4$\\
\tableline
\end{tabular}}
\caption{Parameters used in order to fit the ISO-LWS continuum.}
\label{tab:Paramenters}
\end{center}
\end{table}

The flux errors on the ISO-LWS data are 10$\%$-15$\%$ \citep{Gry03}. Assuming an error of $\sim$12$\%$ for each detector, the goodness of the fit on Figure~\ref{fig:Total_continuum} is $\chi^2$= 0.89 (LW2 flux was not taken into account on the estimation of $\chi^2$). The CND cold component has to be included in the model in order to fit accurately the SPIRE fluxes at wavelengths $\geq250$ $\mu$m. Models with just one CND component can also reproduce the ISO-LWS continuum very precisely. However, the model must satisfy another constraint set by the fluxes of the atomic and molecular lines detected by ISO-LWS in this region. We have used a radiative transfer code as described in \citet{Gonzalez02}, to simultaneously fit the continuum and the spectral lines, obtaining the best results for the case of two CND components. We present the discussion of the gas properties through SgrA$^*$ in a separate paper \citep{White11}.

\section{Conclusions}

We present new Herschel PACS and SPIRE maps of the Galactic Centre, together with previously unpublished ISO-LWS 10-band parallel mode maps that enable us to obtain an accurate calibration for the central parsecs and a detailed spectral shape around the wavelength peak of the continuum emission.  Photometry from the data set allows us to model dust in the central 2 pc of the Galactic Centre. Combined with MSX-E 21.3 $\mu$m images, we use these new results to trace the dust properties throughout the Galactic Centre region. 

The ISO LWS continuum in the direction of Sgr A*, extended with Herschel data out to 500 $\mu$m, is best fit by a model of a central cavity of 1 pc radius emitting like a greybody with T = 90 K and $\beta$= 1.23 and with low density n(H$_2$)= 2.8$\times10^3$ cm$^{-3}$.  The region is surrounded by the circumnuclear disk represented by two different components: a hot component with a dust temperature of 44.5 K and n(H$_2$)= 1.19$\times10^4$ cm$^{-3}$ with $\beta$= 1.35, and a cold component with T$_{dust}$= 23.5 K, n(H$_2$)= 2.41$\times10^4$ cm$^{-3}$ and $\beta$= 1.4. The two components indicate that the CND is not a single uniform body, but (as suggested by the images and previous authors) has numerous clumps, gaps, and substructures but whose average properties manifest these two basic behaviours. This simple model is able to reproduce the observed ISO-LWS continuum as well as the SPIRE data at 250 $\mu$m, 350 $\mu$m and 500 $\mu$m.  The MSX-E map traces the emission of the very small grains population through the Radio Arc, the Sickle, the Arched Filaments and confined in the central cavity in SgrA$^*$. These spatial structures are now directly detected in the far infrared, and, with the full SED available, their dust features are all characterized.  As expected, the largest grains are distributed throughout the CND and correspond to the densest and coolest molecular clouds. Unresolved issues with the Herschel photometry in the current dataset leaves the absolute accuracy of the coldest, densest regions (like G0.11-0.11) approximate.

Dust opacity properties in the central cavity are typical of grains with a thin ice layer; while the dust properties in the CND, with lower temperature and highest densities, is more typical of dust grains with thick ice mantles. 

The total integrated continuum flux around the inner edge of the CND and the central cavity provides a total mass of M$\sim$$4.9 \times 10^4$ M$_{\odot}$. The luminosities obtained are  L$_{cavity}\sim 2.2\times10^6$ $L_{\odot}$ and L$_{CND}\sim 1.5\times10^6$ $L_{\odot}$ in the central $\sim$2.0 pc in SgrA$^*$.
\acknowledgments

We thank the referee for his or her careful reading of the paper, and 
for the useful comments made. We are very grateful to Tanya Lim for her suggestions on the treatment of the ISO LWS parallel mode data. This work was supported in part by NASA Grant NNX10AD83G. The ISO-LWS LIA is a joint development of the ISO-LWS Instrument Team at Rutherford Appleton Laboratory and the Infrared Processing and Analysis Center (IPAC/Caltech, USA). HIPE is a joint development by the Herschel Science Ground Segment Consortium consisting of ESA, the NASA Herschel Science Center, and the HIFI, PACS and SPIRE consortia. 
\notetoeditor{} 


\begin{thebibliography}{}
\bibitem[Bally et al.(2010)]{Bally10}Bally, J., Aguirre, J., Battersby, C., Bradley, E. T., Cyganowski, C., Dowell, D., et al. 2010, \apj, 721, 137
\bibitem[Bernard et al.(2010)]{Bernard10}Bernard, J.-Ph., Paradis, D., Marshall, D.J., Montier, L., et al. 2010, \aap, 158, L88
\bibitem[Boulanger et al.(1985)]{Boulanger85}Boulanger, F., Baud, B., van Albada, G.D. 1985, \aap, 144, L9
\bibitem[Bradford et al.(2005)]{Bradford05}Bradford, C.M., Stacey, G.J., Nikola,T., et al. 2005, \apj, 623, 866
\bibitem[Butler et al.(2009)]{Butler09}Butler, M.J., Tan, J.C. 2009, \apj, 696, 484
\bibitem[Casassus et al.(2008)]{Casassus08}Casassus, S., Dickinson, C., Cleary, K., Paladini, R., Etxaluze, M., Lim, T., White, G. J., Burton, M., Indermuehle, B., Stahl, O., Roche, P. 2008, \mnras, 391, 1075
\bibitem[Chan et al.(2001)]{Chan01}Chan S. J. et al. 2001, in Metcalfe L., Kessler M. F., eds, The Calibration Legacy of the ISO Mission. ESA SP-481
\bibitem[Cohen(1999)]{Cohen99}Cohen, M. 1999, "New Perspectives on the Interstellar Medium", ASP Conferences Series, Vol. 168
\bibitem[Colgan et al.(1996)]{Colgan96}Colgan, S. W. J., Erickson, E. F., Simpson, J. P., Haas, M. R., Morris, M. 1996, \apj, 470, 882
\bibitem[Cotera et al.(1996)]{Cotera96}Cotera, A.S., Erickson, E.G., Colgan, S.W.J., Simpson, J.P., Allen, D.A., Burton, M. 1996, \apj, 461, 750
\bibitem[Desert et al.(1990)]{Desert90}D\'esert, F.X., Boulanger, F., Puget, J.L. 1990, \aap, 237, 215
\bibitem[Dupac et al.(2003)]{Dupac03}Dupac, X., Bernard, J. P., et al. 2003, \aap, 404, L11
\bibitem[Egan et al.(1999)]{Egan99}Egan, M.P., Price, S.D., Moshir, M.M., Cohen, M., Tedesco, E., Murdock, T.L., Zweil, A., Burdick, S., Bonito, N., Gugliotti, G.M., Duszlak, J. 1999, Air Force Research Laboratory Technical Report No. AFRL-VS-TR-1999-1522
\bibitem[Genzel et al.(1982)]{Genzel82}Genzel, R., Watson, D., Townes, C., et al. 1982, in The Galactic center; Proceedings of the Workshop, Pasadena, CA, January 7, 8, 1982 (A83-40676 19-90). New York, American Institute of Physics, p. 72-76
\bibitem[Genzel et al.(1987)]{Genzel87}Genzel, R. and Townes, C.H. 1987, \araa, 25, 377 
\bibitem[Ghez et al.(2008)]{Ghez08}Ghez, A.M., Salim, S., Weinberg, N.N., Lu, J.R., Do, T., Dunn, J.K., Matthews, K., Morris, M.R., Yelda, S., Becklin, E.E., Kremenek, T., Milosavljevic,M. Naiman, J. 2008, \apj, 689, 1044
\bibitem[Gillesen et al.(2009)]{Gillessen09}Gillessen, S., Eisenhauer, F., Trippe, S., Alexander, T., Genzel, R., Martins, F., Ott, T. 2009 \apj, 692, 1075
\bibitem[Gonz\'alez-Alfonso et al.(2002)]{Gonzalez02}Gonz\'alez-Alfonso, E., Wright, C. M., Cernicharo, J., Rosenthal, D., Boonman, A. M., van Dishoeck, E. F. 2002, \aap, 386, 1074
\bibitem[Griffin et al.(2006)]{Griffin06}Griffin, M., Abergel, A., Ade, P., Angr\'e, P., Baluteau, J-P., Bock, J., et al. 2006, SPIE, 6265, 7
\bibitem[Gry et al.(2003)]{Gry03}Gry, C., Swinyard, B., Harwood, A., Trams, N., Leeks, S., Lim, T., Sidher, S., Lloyd, C., Pezzuto, S., Molinari, S., Lorente, R., Caux, E., Polehampton, E., Chan, J., Hutchinson, G., M\"uller, T., Burgdorf, M., Grundy, T. 2003, "The ISO Handbook", SAI-99-077/Dc, Version 2.1
\bibitem[Gusten et al.(1981)]{Gusten81}G\"usten, R., Walmsley, C. M., \& Pauls, T. 1981, \aap, 103, 197
\bibitem[Herrnstein et al.(2005)]{Herrnstein05} Herrnstein, R. M., Ho, P. T. P. 2005, \apj, 620
\bibitem[Hirashita et al.(2007)]{Hirashita07}Hirashita, H., Hibi, T., Shibai, H. 2007, \mnras, 379, 974
\bibitem[Huttemeister et al.(1993)]{Huttemeister93}H\"uttemeister, S., Wilson, T. L., Banina, T. M., \& Mart\'in-Pintado, J. 1993, \aap, 280, 255
\bibitem[Kerr et al.(1986)]{Kerr86}Kerr, F. J. \& Lynden Bell, D. 1986, \mnras, 221, 1023
\bibitem[Lang et al.(2001)]{Lang01}Lang, C.C., Goss, W.M., \& Morris, M. 2001, \aj, 121 
\bibitem[Latvakoski et al.(1999)]{Latvakoski99}Latvakoski, H.M., Stacey, G.J., Gull, G.E., Hayward, T.L. 1999, \apj, 511, 761
\bibitem[Lewis et al.(2000)]{Lewis00}Lewis, G. F., Chapman, S. C. 2000, \mnras, 318, L31-L33
\bibitem[Lim et al.(1999)]{Lim99} Lim, T., Vivares, F., Caux, E. 1999,``The LWS serendipity mode", presented at ``ISO Surveys of dusty universe meeting"
\bibitem[Liu et al.(2011)]{Liu11}Liu, S., Fryer, C.L., Li H. 2011, The Open Astronomy Journal, 4, 38
\bibitem[Lutz et al.(1996)]{Lutz96}Lutz, D., Feuchtgruber, H., Genzel, et al. 1996, \aap, 315, L269
\bibitem[Martin et al.(2004)]{Martin04}Martin, C. L., Walsh, W. M., Xiao, K., Lane, A. P., Walker, C. K., Stark, A. A. 2004, \apj, 150:239 
\bibitem[McGary et al.(2001)]{McGary01}McGary, R., Coil, A. and Ho, P.T.P. 2001, \apj, 559, 326
\bibitem[Moneti et al.(2001)]{Moneti01}Moneti, A., Stolovy, S., Blommaert, J. A. D. L., Figer, D. F., Najarro, F. 2001, \aap, 366, 106
\bibitem[Montero-Casta\~no et al.(2009)]{Montero09}Montero-Casta\~no, M., Herrnstein, R. M. Ho, P. T. P. 2009, \apj, 695
\bibitem[Morris et al.(1995)]{Morris95}Morris, M., Davidson, J. A., Werner, M. W. 1995, "Airborne Astronomy Symposium on the Galactic Ecosystem", ASP Conference Series, Vol. 73 
\bibitem[Nagayama et al.(2009)]{Nagayama09}Nagayama, T., Omodaka, T., Handa, T., Toujima, H., Sofue, Y., Sawada, T., Kobayashi, H., and Koyama, Y. 2009, \pasj, Vol. 61, 1023
\bibitem[Ossenkopf \& Henning (1994)]{Ossenkopf94}Ossenkopf, V., \& Henning, Th. 1994, \aap, 291, 943
\bibitem[Poglitsch et al.(2001)]{Poglitsch01}Poglitsch, A. Waelkens, C., Geis, N. 2001, in "The Promise of the Herschel Space Observatory", ESA SP-460
\bibitem[Pollack et al.(1994)]{Pollack94}Pollack, J. B., Hollenbach, D., Beckwith et al. 1994, \apj, 421, 615
\bibitem[Pomar\'es et al.(2009)]{Pomares09}Pomar\'es, M., Zavagno, A., Deharveng, L., Cunningham, M., Jones, P., Kurtz, S., et al. 2009, \aap, 494, 987
\bibitem[Rodr\'iguez-Fern\'andez et al.(2004)]{Rodriguez04}Rodr\'iguez-Fern\'andez, N. J., Mart\'in-Pintado, J., Fuente, A., Wilson, T. L. 2004, \aap,427, 217 
\bibitem[Shields \& Ferland(1994)]{Shields94}Shields, J.C. \& Ferland, G.J. 1994, \apj, 430, 236
\bibitem[White et al.(2011)]{White11}White, G. J., Etxaluze, M., Smith, H. A., Gonz\'alez-Alfonso, E., Stark, A. A., et al. 2011, \aap, in preparation.
\bibitem[Wright et al.(2001)]{Wright01} Wright, M.C., Coil, A.L., McGary, R., Ho, P.T.P. and Harris, A.I. 2001, \apj., 551, 254
\bibitem[Yusef-Zadeh et al.(1984)]{Yusef84} Yusef-Zadeh, F., Morris, M. \& Chance, D. 1984, Nature, 310, 55
\bibitem[Yusef-Zadeh et al.(1987)]{Yusef87} Yusef-Zadeh, F. \& Morris, M. 1987, \apj, 320, 545
\bibitem[Yusef-Zadeh et al.(2001)]{Yusef01} Yusef-Zadeh, F., Stolovy, S., Burton, M., Wardle, M. and Ashley, M.C.B. 2001, \apj., 560, 749
\bibitem[Yusef-Zadeh et al.(2007)]{Yusef07}Yusef-Zadeh, F., Wardle, M., Roy, S. 2007, \apj., 665, L123 
\bibitem[Yusef-Zadeh et al.(2009)]{Yusef09}Yusef-Zadeh, F., Hewitt, J. W., Arendt, R. G., Whitney, B., Rieke, G., Wardle, M., Hinz, J. L., Stolovy, S., Lang, C. C., Burton, M. G., Ramirez, S. 2009, \apj, 702, 178 
\bibitem[Zylka et al.(1994)]{Zylka94} Zylka, R., Mezger, P. G., Wilson, T. L. et al. 1994, p 161, In the Nuclei of Normal Galaxies, edited by R. Genzel and A. I. Harris. Kluwer Academic Publishers
\bibitem[Zylka et al.(1995)]{Zylka95} Zylka, R., Mezger, P.G., Ward-Thompson, D., Duschl,W.J. and Lesch, H. 1995, \aap, 297, 83
\end{thebibliography}
\end{document}